\documentclass[
twocolumn, 
superscriptaddress,
nofootinbib,
amsmath,amssymb,
longbibliography,
prr,
]{revtex4-2}

\usepackage[utf8]{inputenc}
\usepackage{CJKutf8}
\usepackage{amsmath}
\usepackage{amsfonts}
\usepackage{amssymb}
\usepackage{physics}
\usepackage{color}
\usepackage{graphicx}
\usepackage{dcolumn}
\usepackage{bm}
\usepackage{hyperref}

\newcommand{\ie}{{\it i.e.~}}

\begin{document}

\title{On Quantum Reliability Characterizing Systematic Errors in Quantum Sensing}

\author{Lian-Xiang Cui (\begin{CJK}{UTF8}{gbsn}崔廉相\end{CJK})}
\affiliation{Beijing Computational Science Research Center, Beijing 100193, China}

\author{Yi-Mu Du (\begin{CJK}{UTF8}{gbsn}杜亦牧\end{CJK})}
\email{ymdu@gscaep.ac.cn}
\affiliation{Graduate School of CAEP, Beijing 100193, China}

\author{C.P. Sun (\begin{CJK}{UTF8}{gbsn}孙昌璞\end{CJK})}
\email{suncp@gscaep.ac.cn}
\affiliation{Graduate School of CAEP, Beijing 100193, China}
\affiliation{Beijing Computational Science Research Center, Beijing 100193, China}

\date{\today}

\begin{abstract}
Quantum sensing utilize quantum  effects, such as entanglement and coherence, to measure physical signals. 
The performance of a sensing process is characterized by error which requires comparison to a true value. However, in practice, such a true value might be inaccessible. In this study, we utilize quantum reliability as a metric to evaluate quantum sensor's performance based solely on the apparatus itself, without any prior knowledge of true value. We derive a general relationship among reliability, sensitivity, and systematic error, and demonstrate this relationship using a typical quantum sensing process. That is to measure magnetic fields (as a signal) by a spin-$1/2$ particle and using the Stern-Gerlach apparatus to read out the signal information. Our findings illustrate the application of quantum reliability in quantum sensing, opening new perspectives for reliability analysis in quantum systems.
\end{abstract}

\maketitle

\section{Introduction}

A sensor's functioning
highly relies on its measurement process, which converts  observations into numbers to quantify the phenomena, thereby  shaping the depth and clarity of our understanding of the world. The reliability of these measuring apparatus plays a pivotal role. 
The readings of any apparatus usually consist of the true value of the measured signal and the error introduced during the measurement process \cite{test_theory_2008}, with the performance of the apparatus highly related to these two factors. The discourse on measurement errors holds profound significance across various fields. From the physiological data of patients in medicine \cite{Health_Syst_2008,measurement_reliability_in_clinical_trials_2004} to the experimental data in biochemistry laboratories \cite{hibbert2007}, from the minutiae of atomic frequency \cite{Quantu_sensing_RevModPhys} to the grand detection of gravitational waves in the cosmos \cite{gravitational_wave_detectors_2012, LIGO_PhysRevLett_2019}, the accuracy of measurement is of utmost importance. Meanwhile, the field of reliability engineering also places significant emphasis on the measurement of errors \cite{measurement_uncertainties_RESS_2023, Accelerated_degradation_testing_RESS_2024,gamma_process_RESS_2016,Bayesian_measurements_approach_RESS_1997, degradation-based_burn-in_RESS_2016}.

Recently, quantum sensing, as a new technology on highly precise measurements, are widely concerned \cite{Quantu_sensing_RevModPhys, Quantum_metrology_RevModPhys, nondemolition_measurements_RevModPhys_1996, Search_for_new_physics_RevModPhys_2018}. 
It encompasses the utilization of quantum systems or effects to measure physical quantities, whether they be classical or quantum, such as electromagnetic fields, time or frequency, temperature, and pressure. Quantum measurement apparatus exploit a key characteristic of quantum systems: their heightened sensitivity to external perturbations, which enables highly precise measurements. Additionally, intrinsic quantum properties like coherence and entanglement \cite{Griffiths_Quantum_Mechanics, Weinberg_Lectures_on_Quantum_Mechanics, Quantum_entanglement_RevModPhys} are harnessed to perform these measurements. Many quantum measurement apparatus have already been integrated into our daily lives, including nuclear magnetic resonance\cite{Rabi_PhysRev_1938,basics_of_nuclear_magnetic_resonance_2008,Magnetoencephalography_RevModPhys_1993, quantum_diamond_sensors_RevModPhys_2024} and atomic clocks\cite{atomic_clocks_RevModPhys}.

A fundamental aspect of evaluating the performance of a measurement apparatus is the assessment of error. Typically, assessing the systematic error of such an apparatus requires a true value of the measured quantity for comparison. This true value is ideally provided by a "yardstick"-a perfectly standard apparatus. However, when the true value is inaccessible, an alternative metric must be identified to characterize the apparatus's performance. In such cases, reliability emerges as a crucial indicator.

Reliability measures an apparatus's ability to consistently perform its function. It is determined by the performance of each component in the system; as long as each step operates as designed, the output remains reliable. Importantly, in measurement processes, reliability is intrinsic to the apparatus and does not depend on the true value being measured. Unlike error, which focuses on the outcome, reliability takes a reductionist approach by evaluating the quality of each individual step in the process. Thus, while error is concerned with results, reliability concentrates on the process.

In the current study, we characterize the measurement error by reliability.
A relationship among reliability, sensitivity, and systematic error of measurement apparatus is derived, showing that when the apparatus is near ideal, there is a proportional relationship between the decrease in reliability and the sensitivity times systematic error, \ie
\begin{equation}
    1 - \text{Reliability} \sim \text{Sensitivity} \times \text{Error}.
    \label{eq:R and S and E}
\end{equation}
Here, the proportionality coefficient is determined solely by the measuring apparatus and is independent of the physical quantity being measured. From this relationship, we can infer the systematic error of the apparatus through its reliability, thereby eliminating the need for a standard reference scale. We then focus on quantum sensing process, using a typical quantum sensing model to illustrate this relationship: sensing a magnetic field via spin and employing the Stern-Gerlach (SG) apparatus to measure the spin state.

This paper is structured as follows. Section II provides a brief introduction to the concepts of measurement error and derives the relationship among reliability, sensitivity, and systematic error. Section III explores sensing and reliability within the quantum scheme. Section IV illustrates the relationship discussed in Section II through a specific quantum sensing process. Finally, Section V summarizes our findings and suggests directions for future research.

\section{Measurement error and reliability}
Measurement is the process of converting a physical quantity into a numerical reading. This conversion can be direct, such as determining length using a ruler, or indirect, such as transforming a signal into a reading via sensors. Here, we refer to the entire progression from the physical quantity to its numerical representation as the measurement process, which is the focus of our study in this paper. Before commencing our discussion, it is essential to define measurement errors clearly. This section specifically delineates systematic and random errors and derives the relationship among the reliability, sensitivity, and systematic errors of measuring apparatus.

\subsection{Measurement error}
Measurement has long been a topic of significant concern. When describing measurements, it is imperative to carefully define the terminology used to ensure a common foundational basis for discussion. Numerous international organizations, including the Bureau International des Poids et Mesures (BIPM), the International Electrotechnical Commission (IEC), the International Organization for Standardization (ISO), and the International Organization of Legal Metrology (OIML), etc., have come together to publish manuals that precisely defines these terms, International Vocabulary of Metrology (VIM) \cite{VIM} and Guide to the Expression of Uncertainty in Measurement (GUM) \cite{GUM}.

\begin{figure}[htbp]
    \centering
    \includegraphics[width=0.90\linewidth]{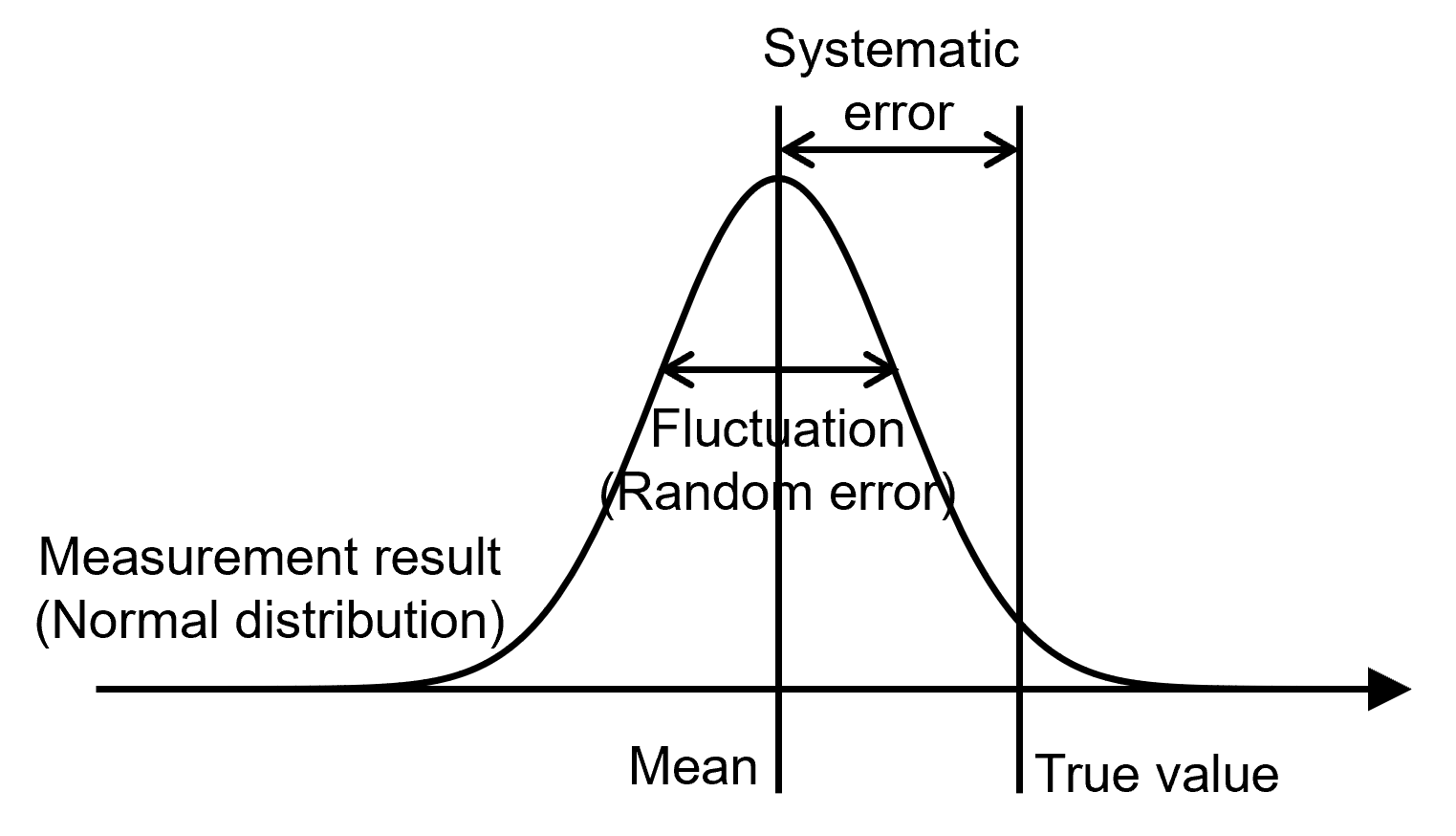}
    \caption{Schematic Representation of Measurement Error.}
    \label{fig:Measurement Error}
\end{figure}

Here, we focus on measurement error, a measured quantity value minus a reference quantity value (true value). Measurement errors are traditionally categorized into two types: random errors and systematic errors \cite{hibbert2007}.
Let \( x \) represent the true value of the measurand and \( \hat{x} \) the observed measurement. The observed measurement can then be expressed as the sum of the true value $x$, random error $\epsilon$, and systematic error $\delta$, \ie, 
\begin{equation}
    \hat{x} = x  + \epsilon + \delta.
\end{equation}
Random error $\epsilon$ refers to the error that in replicate measurements varies unpredictably.
The results of multiple measurements typically yield a distribution, which, under normal circumstances, follows a Gaussian distribution. The width of this Gaussian distribution, characterized by its variance (fluctuations), is attributed to random noise. In contrast, the deviation of the mean of this distribution from the true value is due to systematic error $\delta$, as illustrated in Fig.~\ref{fig:Measurement Error}.

It is important to note that the aforementioned errors rely on a true value, which serves as a reference quantity. In the absence of access to this true value in practical scenarios, does this imply an inability to assess the apparatus's quality? In the subsequent discussion, we will evaluate the reliability of the measurement apparatus to ascertain the accuracy of its readings. Specifically, if the apparatus operates as designed, \ie,it operates reliably, its readings should be accurate.

\subsection{Reliability, sensitivity and systematic error}
Reliability is defined as the probability that a product, system, or service will perform its intended function adequately for a specified period of time, or will operate in a defined environment without failure \cite{reliability_2024}. The function of a measurement apparatus is to provide accurate values corresponding to the measurand. Therefore, we can assess the magnitude of an apparatus's error by evaluating its reliability. 

Reliability can be determined by a set of intrinsic parameters that are independent of external information. However, in the context of measurement apparatus, an intriguing question arises: what is the relationship between their reliability and measurement errors? Here, we present an analysis of the relationship among reliability, sensitivity, and systematic errors of measurement apparatus, as shown in Eq.~\eqref{eq:R and S and E}.
This relationship holds when the apparatus is nearly ideal, whether in classical or quantum systems. However, in quantum systems, the definition of reliability differs, as detailed in Section \ref{sec:quantum_measurement}.

\begin{figure}[htbp]
    \centering
    \includegraphics[width=0.95\linewidth]{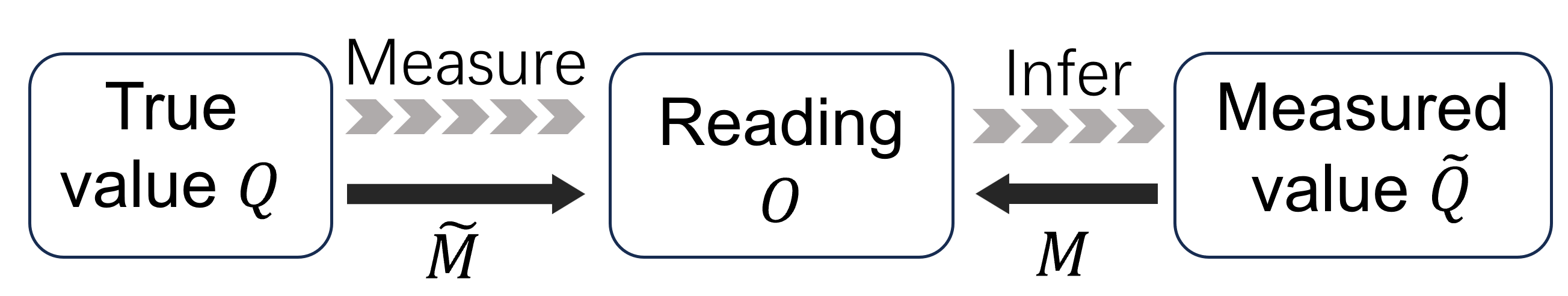}
    \caption{Diagrammatic description of definitions.}
    \label{fig:def of quantities}
\end{figure}

To derive the aforementioned relationship, we denote the true value of the physical quantity to be measured as \( Q \), and the reading obtained from the measurement apparatus as \( O \). Denote the measured value of the target physical quantity, derived from the apparatus reading, as \( \tilde{Q} \).
We define two mappings here. First one is the mapping $\tilde M$, from the true value of the physical quantity \( Q \) to the reading of the measurement apparatus \( O \), \ie  $\tilde M(Q)=O$. Second one is the mapping $M$, from the measured value \( \tilde{Q} \) to the reading \( O \), \ie $M(\tilde Q)=O$. The definitions are illustrated in Fig.~\ref{fig:def of quantities}.

The sensitivity $S$ of the measurement apparatus is defined as the variation in the apparatus's reading in response to a change in the signal being measured, expressed as follows:
\begin{equation}
   S \equiv \frac{{\rm d} \tilde{M}}{{\rm d} Q}.
\end{equation}
The systematic error defined as
\begin{equation}
    \delta Q = \tilde{Q}-Q,
\end{equation} 
the difference of the measured value $\tilde{Q}$ and the true value $Q$.

When the apparatus operation is nearly perfect, there exists a small discrepancy between real and ideal measurement processes. The linear perturbation  implies that an arbitrary metrics on reliability $R$ (where the reliability of the ideal system is normalized to $1$) that quantifies this discrepancy \(1 - R\), should be proportional to measurement variation $\delta M(Q)$. Formally, this relationship is given by:
\begin{equation}
    1 - R \sim \delta M(Q).
    \label{eq:linear approx}
\end{equation}
The variation \(\delta M(Q)\) can be expressed through a series expansion as:
\begin{equation}
    \delta M(Q)= \frac{{\rm d} M}{{\rm d} Q} \delta Q + \frac{1}{2} \frac{{\rm d^2} M}{{\rm d} Q^2} \delta^2 Q + \cdots.
\label{eq:M_expansion}
\end{equation}
When the first-order term is predominant, we obtain the simplified relation:
\begin{equation}
    1 - R \sim   \frac{{\rm d} M}{{\rm d} Q} \delta Q.
\end{equation}
Applying definitions of sensitivity and error, this leads to the relationship found in Eq.~\eqref{eq:R and S and E}.

In Section \ref{sec:example}, we will validate this relationship through a quantum sensing model. However, it is essential that we first clarify certain quantum concepts, which will be addressed in the following section.

\section{Quantum sensing and quantum reliability}
\label{sec:quantum_measurement}
In the quantum realm, the concepts of measurement and reliability both differ significantly from those in the classical context. In quantum mechanics, quantum measurement generally refers to the measurement of quantum states. The comprehensive process we discussed earlier is known as quantum sensing, which will be the focus of our analysis. In this section, we will explore the concepts of quantum sensing and quantum reliability in detail.

\subsection{Quantum sensing}
Quantum sensing refers to the use of quantum systems or quantum effects to assess physical quantities (classical or quantum) \cite{Quantu_sensing_RevModPhys}. However, due to the unique characteristics of quantum mechanics-such as Born rule and the inherent uncertainty principle-it is often infeasible to directly measure the physical quantities of interest. A quantum sensing process generally encompasses several fundamental stages: the initialization of the quantum sensor, its interaction with the signal of interest, and the readout of the final state.

\begin{figure}[htbp]
    \centering
    \includegraphics[width=0.95\linewidth]{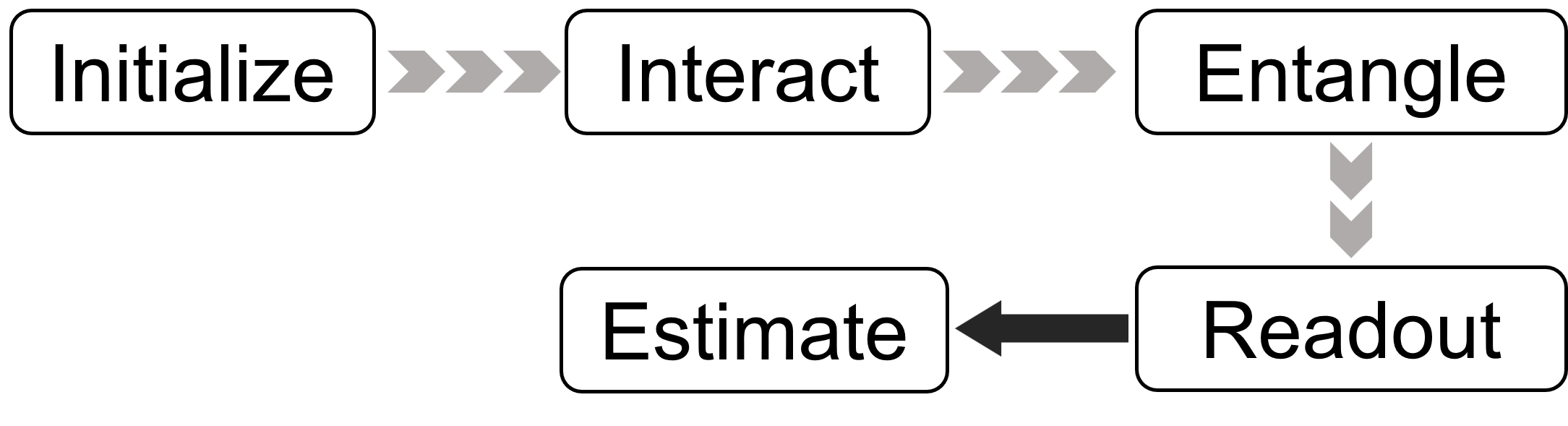}
    \caption{Schematic Diagram of the Quantum Measurement Process. The initial four steps are quantum processes, represented by grey separated arrows. The final step involves the estimation of the signal of interest through readout, which is a classical process, depicted by black arrows.}
    \label{fig:Quantum_Measurement}
\end{figure}

A quantum sensor is a discrete energy level system, typically modeled as a two-level system \cite{divincenzo2000}. Quantum sensing involves inferring the properties of a target signal by examining its impact on the sensor. The procedure is as follows: initially, the sensor is prepared in a specified initial state and interacts with the target signal over a designated period. Subsequently, the quantum state of the sensor is entangled with the the measurement basis, a step commonly known as quantum premeasurement. Projective measurements are then performed to the measurement basis, which often requires either repeating the process multiple times or averaging over an ensemble to determine the state. These readings allow for the inference of the target signal's characteristics. The entire process is illustrated in Fig.~\ref{fig:Quantum_Measurement}, with further details in Ref.~\cite{Quantu_sensing_RevModPhys}.

In contrast to classical measurement, quantum sensing processes are inherently complex, with each step introducing potential errors. Consequently, assessing the reliability of quantum sensing necessitates careful consideration of the operational details at every stage. In this context, employing a process metric to characterize the overall system reliability offers more comprehensive insights than merely comparing result discrepancies. This approach facilitates a more precise discussion of errors. Quantum reliability, as a process metric, accurately describes whether the system's state evolves according to the intended design at each step. In the following, we introduce the concept of quantum reliability.

\subsection{Quantum reliability}
Quantum reliability, introduced by Cui, Du, and Sun \cite{QuantumReliability}, provides a novel metric for assessing the reliability of a quantum system. They define the reliable state of quantum systems and quantify the probability of reliable evolution through the consistent quantum theory \cite{griffiths1984consistent}, a measure termed as quantum reliability. We briefly introduce the concepts here.

The state of a quantum system is described by a wave function \(\ket{\psi} \in \mathcal{H}\), where \(\mathcal{H}\) is the state space of the system, known as the Hilbert space. The properties of the system can be characterized using projection operators \(E\) within this Hilbert space. If the state \(\ket{\psi}\) resides in the subspace \(\mathcal{E}\) corresponding to \(E\), i.e., \(E\ket{\psi} = \ket{\psi}\), then we say that the state \(\ket{\psi}\) possesses the property \(E\). Analogous to the way that the reliability of a classical system can be described by an indicator, the reliability of a quantum system can be characterized by a projector. This allows us to determine whether a quantum system is in a reliable state.

In the preceding analysis, the reliable state of a quantum system is defined. System reliability is inherently a dynamic process. To characterize the system as reliable over a given time interval, the system must maintain its reliable state throughout that interval. Only then can the system's lifetime be defined. The system may undergo a variety of evolutionary processes from time $t_1$ to $t_n$, such as $ \text{reliable}  \to \text{unreliable} \to \text{reliable}\to \cdots \text{reliable}$. Such a sequence of states is denoted as the system's history \( \mathcal{Y} = E_1 \otimes E^\perp_2 \otimes E_3 \otimes \cdots E_n\), where $E^\perp$ is the failure projection of the system and the subscript denotes time. One can define the weight of this history as 
\begin{equation}
\begin{aligned}
    W(\mathcal{Y})=\text{Tr}[E_nU_n\cdots E_2^\perp U_2 E_1 U_1\ketbra{\psi_0}\\
    U_1^\dagger E_1 U_2^\dagger E_2^\perp\cdots U_{n}^\dagger E_n ],
\end{aligned}
\end{equation}
where $\ket{\psi_0}$ is the initial state of the system, $U_i$ is the evolution operator from time $t_{i-1}$ to $t_i$, and $U^\dagger$ is the hermitian conjugate of $U$. To note that the weights associated with these histories do not represent their probabilities. We will subsequently discuss the conditions under which these weights can be interpreted as probabilities.

In reliability analysis, our primary concern is identifying when the system fails; the state of the system after failure is not of interest. Consequently, the histories we are interested in can be represented by the following family $\Theta$:
\begin{equation}
    \begin{aligned}
        \mathcal{F}_1=\ketbra{\psi_0}& \otimes E_1^\perp \otimes I \otimes \cdots I\\
        \mathcal{F}_2=\ketbra{\psi_0} &\otimes E_1\otimes E_2^\perp \otimes I\otimes \cdots I\\
       & \vdots\\
        \mathcal{F}_n=\ketbra{\psi_0}& \otimes E_1\otimes E_2 \otimes \cdots  E_{n-1} \otimes E_n^\perp \\
        \mathcal{R}_n=\ketbra{\psi_0} &\otimes E_1\otimes E_2 \otimes \cdots  E_{n-1} \otimes E_n
    \end{aligned}
\end{equation}
Here, $I$ is the identity, $\mathcal{F}_i$ are the failure histories and $\mathcal{R}_n$ is the survival history.

Define the inner product of two histories as
\begin{equation}
\begin{aligned}
        \left< \mathcal{Y}^\alpha , \mathcal{Y}^\beta \right>
    \equiv
    \text{Tr}[Y^\alpha_n U_n Y^\alpha_{n-1}\cdots Y^\alpha_2 U_2 Y^\alpha_1 U_1\ketbra{\psi_0}\\
    U_1^\dagger Y^\beta_1 U_2^\dagger Y^\beta_2\cdots U_{n}^\dagger Y^\beta_n ],
\end{aligned}
\end{equation}
where  $\mathcal{Y}^k = \ketbra{\psi_0} \otimes Y^k_1\otimes Y^k_2\otimes \cdots Y^k_n, \ k=\alpha ,\beta $.
One can see that, the weight of a history $\mathcal{Y}$ is  $W(\mathcal{Y})=\left<\mathcal{Y},\mathcal{Y} \right>$. 
When the given family satisfies the consistency condition \cite{griffiths1984consistent, Griffiths_2001}, the weights of the histories can be interpreted as the probabilities of these histories. The consistency condition reads:
In a given family $\Theta$, if the inner product of any two distinct histories is zero, then the family is consistent, \ie,
\begin{equation}
    \left< \mathcal{Y}^\alpha , \mathcal{Y}^\beta \right> =0\ \text{for}\  \forall \mathcal{Y}^\alpha \neq \mathcal{Y}^\beta \in \Theta.
\end{equation}

Thus, the reliability of a quantum system at time \( t_n \) as the probability of its survival history, denoted \( R(t_n) = W(\mathcal{R}_n) \). The probability that the system has a lifetime of \( t_k \) is given by \( W(\mathcal{F}_k) \), \ie, the probability that the system fails until time $t_k$.

Recalling Fig.~\ref{fig:Quantum_Measurement}, one could define quantum reliability with setting $E_1$ and $E_2$ after the "interact" and the "Entangle"  step, receptively. The former is to check whether the signal has been successfully encoding into the sensor's state and the latter is to check whether the encoded state has been successfully correlated with  the measurement apparatus's pointer state.

\section{Measuring magnetic field strength with Stern-Gerlach apparatus} 
\label{sec:example}

\begin{figure}[htbp]
    \centering
    \includegraphics[width=0.95\linewidth]{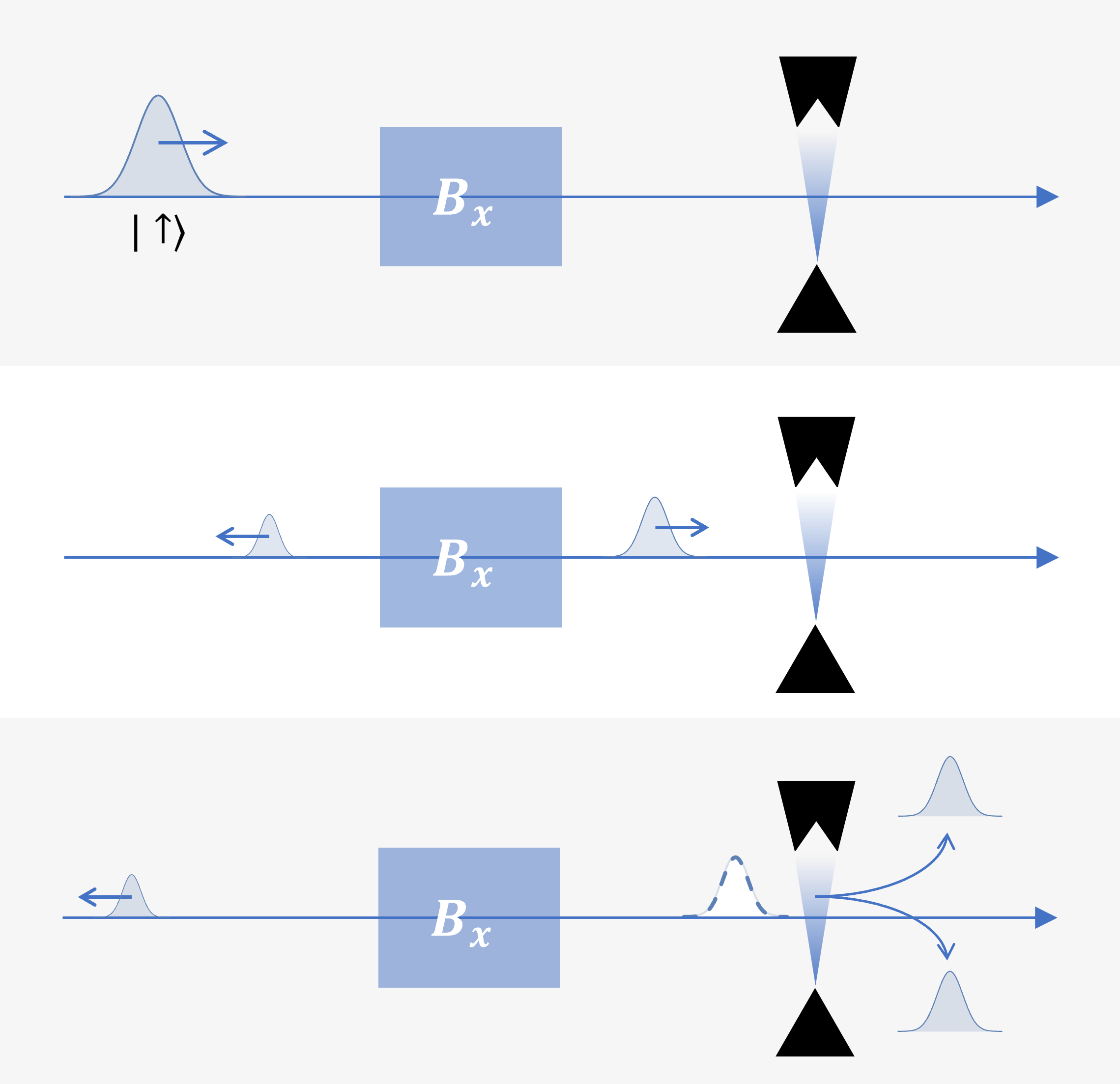}
    \caption{Schematic diagram of the quantum sensing process under consideration.}
    \label{fig:Schematic_Diagram}
\end{figure}

In this section, we demonstrate the relationship outlined in Eq.~\eqref{eq:R and S and E} through a quantum sensing process, utilizing a spin-1/2 particle and a Stern-Gerlach apparatus to detect a magnetic field. Here, the spin-1/2 particle acts as a quantum sensor, with its energy level configuration being modified upon interaction with the target magnetic field. Specifically, consider a spin-1/2 particle initially in the $+1$ eigenstate of $\sigma_z$, denoted as $\ket{\uparrow}$. This particle passes through a magnetic field $B_x$ oriented along the $x$-direction, causing the spin to precess. A SG apparatus is then employed to measure the populations of the spin-up and spin-down states. From these measurements, the magnitude of the magnetic field can be inferred. The schematic diagram of this setup is shown in Fig. \ref{fig:Schematic_Diagram}.

\subsection{Ideal process}
The initial momentum of the spin is $k_0>0$, and the width of the magnetic field to be measured is $a$. The evolution of the spin within the magnetic field is governed by the Hamiltonian $H_0=B_x \sigma_x$, where $\sigma_x$ is the Pauli-X matrix. After a time 
$t_1=am/k_0$, the spin state evolves into
\begin{equation}
    \begin{aligned}
        \ket{\psi(t)}&=e^{-i H_0 t_1 / \hbar }\ket{\uparrow}\\
        &=\cos \frac{B_x a m}{k_0 \hbar }\ket{{\uparrow}}-i \sin \frac{B_x a m}{k_0 \hbar} \ket{\downarrow},
    \end{aligned}
\end{equation}
where $m$ is the mass of spin. The population of spin-up and spin-down are
\begin{equation}
    \alpha = \cos^2 \frac{B_x a m}{k_0 \hbar } \text{ and } \beta=\sin^2 \frac{B_x a m}{k_0 \hbar },
\end{equation}
respectively. From this, the magnitude of the magnetic field to be measured can be determined through
\begin{equation}
    B_x = \frac{k_0 \hbar}{a m}\arctan \sqrt{\frac{\beta}{\alpha}}.
    \label{eq:B_x}
\end{equation}

\subsection{Real evolution}
However, in a realistic physical scenario, the incidence of spins into the magnetic field induces scattering, and the SG apparatus cannot perfectly separate the spin-up and spin-down states. This inevitably introduces systematic errors into the measurement. The real Hamiltonian writes
\begin{equation}\label{15}
    H=\frac{p^2}{2m} + \Sigma_1(\mathbf{r}) B_x \sigma_x - \Sigma_2(\mathbf{r})B(z)\sigma_z,
\end{equation}
where $\Sigma_1$ and $\Sigma_2$ are indicator functions that describe the regions influenced by the magnetic field and the SG apparatus, respectively. And $\sigma_x$ and $\sigma_z$ are Pauli-X and Z operator, respectively. 

The initial state in both the $x$ and $z$ directions is a Gaussian wave packet with momentum $k_0$ along $+x$ direction, and the initial state of the spin is the $+1$ eigenstate  $\ket{\uparrow }$ of $\sigma_z$. The total initial state writes
\begin{equation}
    \ket{\Psi_0}=\ket{\uparrow}\otimes \ket{\psi_{x0}} \otimes \ket{\psi_{z0}},
\end{equation}
where 
\begin{equation}
\begin{aligned}
        \psi_{x0}(x)=\braket{x}{\psi_{x0}}&=\left(\frac{1}{2\pi \sigma ^2}\right)^{1/4}e^{ik_0(x-x_0)-\left(\frac{x-x_0}{2\sigma}\right)^2},\\
        \braket{z}{\psi_{z0}}&=\left(\frac{1}{2\pi \sigma ^2}\right)^{1/4}\exp\left[-\left(\frac{z}{2\sigma}\right)^2 \right].
\end{aligned}
\end{equation}
Here $x_0$ and $\sigma$ are the initial position and width of the wave packet, respectively.

Following the analysis framework for quantum reliability, we first define reliable projections. For the region of the target magnetic field, reliability indicates that the spin passes through the field without reflection, achieving complete transmission. The corresponding projection operator is expressed as:
\begin{equation}
    E_1 = \int_a ^\infty I_s\otimes \ketbra{x}\otimes I_z \ {\rm d }x,
\end{equation}
where $I_s$ and $I_z$ are identities in spin space and position-$z$ space, respectively.
For the SG apparatus, reliability signifies the complete spatial separation of spin-up and spin-down components along the \(z\)-axis, represented by the projection 
\begin{equation}
\begin{aligned}
    E_2 &= \ketbra{\uparrow}\otimes \int_b^\infty \ketbra{x} \, {\rm d}x \otimes \int_0^\infty \ketbra{z} \, {\rm d }z\\
    &+\ketbra{\downarrow}\otimes \int_b^\infty \ketbra{x} \, {\rm d}x \otimes \int_{-\infty}^0 \ketbra{z} \, {\rm d }z,
\end{aligned}
\label{Eq:SG Realibality projection}
\end{equation}
where $b$ is the length of the SG apparatus in $x$-direction. Sometimes, the overlap integral between the spatially separated upper and lower wave packets is used to characterize the effectiveness of a SG experiment. The definition of $E_2$ we employ here is equivalent to the description using the overlap integral, as detailed in the derivation provided in Appendix \ref{Appendix:overlap integral}.

Disregarding the multiple reflections of the wave packet, the family formed by these projectors is consistent. The reliability of the system writes
\begin{equation}
    R={\rm Tr}[E_2 U_2 E_1 U_1 \ketbra{\Psi_0} U_1^\dagger E_1 U_2^\dagger E_2 ],
    \label{eq:reliability def}
\end{equation}
where $U_1$ and $U_2$ are the evolution operators corresponding to the spin's passage through the magnetic field and the SG region, respectively.

Notably, the reliable projections $E_1$ and $E_2$ defined here are entirely independent of the magnetic field being measured. That is, these reliable projections do not contain any information regarding the physical quantity being assessed. Instead, they are solely related to the apparatus itself. By using these projections to characterize the apparatus's reliability, we can avoid the need for an external reference value to compare with the apparatus's readings, thereby eliminating the need for a standard reference scale.

Assuming that the target magnetic field and the SG region are sufficiently separated, one can independently calculate the evolution of the two processes: (1) scattering in the magnetic field and (2) evolution in the SG region. This evokes a surrogate model for the system described by Eq.~\ref{15} as follows.

\subsubsection{Scattering in magnetic field}
Consider a one-dimensional wave packet scattering problem in the far-field regime, as sketched in step 2 of Fig.~\ref{fig:Schematic_Diagram}. The Hamiltonian is given by
\begin{equation}
    H_1=\frac{p^2}{2m}+\Sigma_1(x) B_x \sigma_x.
\end{equation}
It can be seen that the magnetic field only affects the spin and the coordinate in the $x$-space, while the wave function in the $z$-direction just diffuses. By decomposing the initial spin into $|\uparrow \rangle =(|+\rangle+|-\rangle)/\sqrt{2}$, we can discuss it in terms of potential barriers$(+)$ and wells$(-)$, that is,
\begin{equation}
    H_1^{(\pm)}=\frac{p^2}{2m} \pm \Sigma_1(x) B_x.
\end{equation}
For $\ket{\pm}$, the corresponding transmission and reflection coefficients can be determined as follows:
\begin{equation}
    \begin{aligned}
        T^{(\pm)} =&\frac{4k_0q_{(+)}e^{i(q_{(\pm)}-k_0)a}}{(k_0+q_{(\pm)})^2-(k_0-q_{(\pm)})^2e^{2iq_{(\pm)}a}},\\
R^{(\pm)} =&\frac{(k_0^2 -q_{(\pm)}^2)(1-e^{2iq_{(\pm)}a})}{(k_0+q_{(\pm)})^2-(k_0-q_{(\pm)})^2 e^{2iq_{(\pm)}a}},
    \end{aligned}
\label{eq:T and R}
\end{equation}
where $q_{(\pm)}=\sqrt{2m(E\mp B_x)/\hbar^2}$ and $E=k_0^2/2m$.

The state after evolution under the Hamiltonian \(H_1\) becomes:
\begin{equation}
\begin{aligned}
    &U_1 \ket{\Psi_0} =e^{-i H_1 t_1 / \hbar } \ket{\Psi_0} =\\
    &\begin{cases}
    \frac{1}{\sqrt{2}} \left(R^{(+)} \ket{+} + R^{(-)} \ket{-} \right)\otimes \ket{\psi_{x0,-}}\otimes \ket{\psi_{z1}}, & x<0\\
    \frac{1}{\sqrt{2}} \left(T^{(+)} \ket{+} + T^{(-)} \ket{-} \right)\otimes \ket{\psi_{x0,+}}\otimes \ket{\psi_{z1}}, & a<x\\
\end{cases}.
\end{aligned}
\label{eq:U1Psi0}
\end{equation}
Here, \(\ket{\psi_{x0,\pm}}\) represents Gaussian wave packets propagating in both positive and negative directions, as defined in Appendix \ref{Appendix:1d scattering}. The wave function in the z-direction has propagated for a time \( t_1 \), resulting in a Gaussian wave packet $\psi_{z1}$ with a width of \(\sigma' = \sigma^2(1+i\hbar t_1 /m\sigma^2 )\). The detailed derivations of Eq.~\eqref{eq:T and R} and Eq.~\eqref{eq:U1Psi0} can be found in Appendix \ref{Appendix:1d scattering}.

\subsubsection{Evolution in  SG apparatus}
The SG apparatus spatially separates particles based on their spin orientation using a non-uniform magnetic field. Here, a SG apparatus with a non-uniform magnetic field along the $z$-axis is considered, as illustrated in step 3 of Fig.~\ref{fig:Schematic_Diagram}. This setup effectively separates particles into spin-up and spin-down orientations along the $z$-axis.

The Hamiltonian of SG apparatus is written with the linear approximation 
\begin{equation}
    H_2 = \frac{p^2}{2m} - B(z)\sigma_z \approx \frac{p^2}{2m} - f z \sigma_z.
\end{equation}
Denote the input state of the SG apparatus as
\begin{equation}
    \ket{\Psi_1}=(c_1\ket{\uparrow} + c_2 \ket{\downarrow})\otimes \ket{\psi_{z1}}  \otimes  \ket{\psi_{x1}} .
    \label{eq:Psi_1}
\end{equation}
After evolving for a time \( t_2 \) in the SG apparatus, the wave function become
\begin{equation}
\begin{aligned}
    U_2&\ket{\Psi_1} = e^{-i H_2 t_2 / \hbar }\ket{\Psi_1}\\
    &=\left(c_1\ket{\uparrow} \otimes \ket{\psi_{z2}^{(+)}} + c_2\ket{\downarrow} \otimes \ket{\psi_{z2}^{(-)}}\right)\otimes \ket{\psi_{x2}},
\end{aligned}
\end{equation}
where
\begin{equation}
\begin{aligned}
        &\psi_{z2}^{(\pm)}(z,t_2)=  \braket{z}{\psi_{z2}^{(\pm)}(t_2)}=\\
    &\frac{\left(\frac{\sigma'^2 }{2\pi}\right)^{1/4}}{\sqrt{\sigma'^2 + \frac{i\hbar t_2}{2m}}} \exp\left[
        -\frac{if^2 t_2^3 }{6\hbar m}-\frac{\left(z\mp \frac{ft_2^2}{2m}\right)^2 }{4\left(\sigma'^2 +\frac{i\hbar t_2}{2m}\right)}\pm ifzt_2/\hbar 
    \right],
\end{aligned}
\end{equation}
and \(\ket{\psi_{x2}}\) is the result of the free evolution of \(\ket{\psi_{x1}}\). This can be obtained through Wei-Norman theory, \ie,
\begin{equation}
\begin{aligned}
    &\exp [{-i(\frac{p^2}{2m}\pm fz)t/\hbar}]\\
    &=\exp[{-\frac{itp^2 }{2m\hbar}\mp\frac{ift^2 p}{2m\hbar}}] \exp[{\mp\frac{iftz}{\hbar}-\frac z{if^2t^3}{6m\hbar}}]. 
\end{aligned}
\end{equation}

Therefore, the populations in the upper half-plane $\tilde \alpha$ and the lower half-plane $\tilde \beta$ of the SG apparatus are
\begin{equation}
    \begin{aligned}
        \tilde{\alpha} = \int_0^\infty \left( |c_1 \psi_{z2}^{(+)}|^2  + |c_2 \psi_{z2}^{(-)}|^2\right) {\rm d} z,\\
        \tilde{\beta} = \int_{-\infty}^0 \left( |c_1 \psi_{z2}^{(+)}|^2  + |c_2 \psi_{z2}^{(-)}|^2\right) {\rm d} z,
    \end{aligned}
\end{equation}
respectively, representing the apparatus's readings. From these readings, we can infer the magnitude of the magnetic field from Eq.~\eqref{eq:B_x}, \ie
\begin{equation}
\tilde{B} = \frac{k_0 \hbar}{a m}\arctan \sqrt{\frac{\tilde \beta}{\tilde \alpha}}.
\end{equation}

\subsection{Reliability of the system}
After analyzing the two evolution processes separately, the system's reliability can be determined using Eq.~\eqref{eq:reliability def}. It is important to note that the incident state in the SG region is given by 
\begin{equation}
\begin{aligned}
    &\ket{\Psi_1} = E_1U_1\ket{\Psi_0}\\
    &=\frac{1}{\sqrt{2}} \left(T^{(+)} \ket{+} + T^{(-)}\ket{-} \right) \otimes \ket{\psi_{x0,+}} \otimes \ket{\psi_{z1}}.
\end{aligned}
\end{equation}
Hence, the coefficients in Eq.~\eqref{eq:Psi_1} read
\begin{equation}
    \begin{aligned}
        c_1=\frac{1}{2}\left(T^{(+)} + T^{(-)}\right),\  c_2 =\frac{1}{2}\left(T^{(+)} - T^{(-)}\right),
    \end{aligned}
\end{equation}
and $\ket{\psi_{x1}}=\ket{\psi_{x0,+}}$.
The state after projection of $E_2$ is 
\begin{equation}
\begin{aligned}
        \ket{\Psi_2} &\equiv E_2U_2\ket{\Psi_1}\\
    &= \int_0^\infty c_1\psi_{z2}^{(+)}(z)\ket{\uparrow} \otimes \ket{z} {\rm d}z \otimes \braket{x}{\psi_{x2}}\ket{x} {\rm d} x\\
    &+ \int_{-\infty}^0 c_2\psi_{z2}^{(-)}(z)\ket{\downarrow} \otimes \ket{z} {\rm d}z
    \otimes \braket{x}{\psi_{x2}}\ket{x} {\rm d} x
\end{aligned}
\end{equation}
The reliability of the system is
\begin{equation}
\begin{aligned}
        R&={\rm Tr} \left[\ketbra{\Psi_2} \right]\\
    &=\int_0^\infty |c_1 \psi_{z2}^{(+)}(z)|^2{\rm d}z +\int_{-\infty}^0 |c_2 \psi_{z2}^{(-)}(z)|^2{\rm d}z.
\end{aligned}
\end{equation}
\begin{figure}[htbp]
    \centering
    \includegraphics[width=0.95\linewidth]{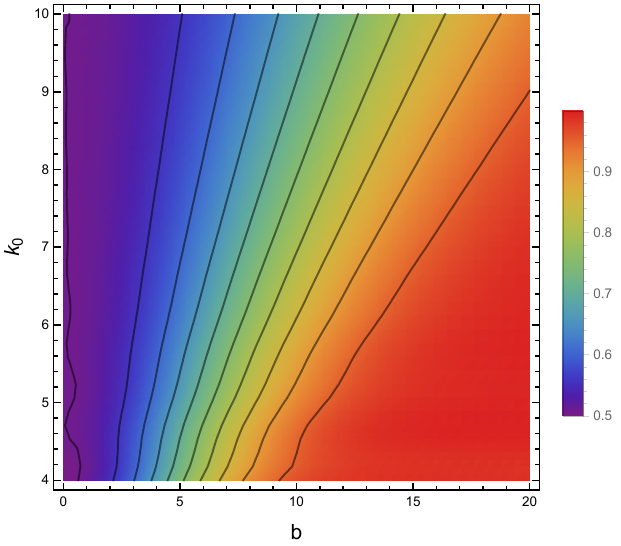}
    \caption{Reliability of the system. The black lines represent contour lines, and their curvature is attributed to resonant tunneling occurring during the scattering process. Parameters: \(\hbar = 1\), \(a = 3\), \(m = 1\), \(\sigma = 0.5\), \(f = 1\), \(B_x = 2\).}
    \label{fig:reliability}
\end{figure}

\begin{figure}[htbp]
    \centering
    \includegraphics[width=0.95\linewidth]{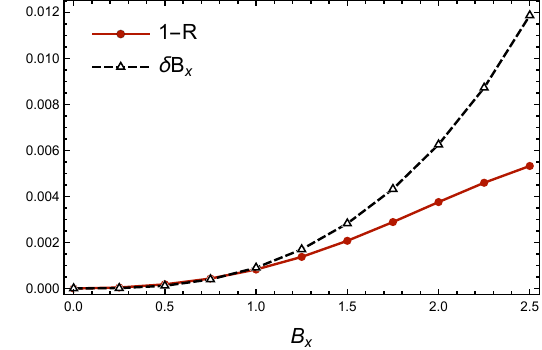}
    \caption{Reliability and Error. $b=35$, $k_0=5$.}
    \label{fig:error}
\end{figure}

Given the magnetic field to be measured, there are two adjustable parameters in the apparatus: the incident momentum \( k_0 \) of the wave packet and the length \( b \) of the SG region. These parameters determine the evolution times of the wave packet in the magnetic field and SG apparatus, denoted as \( t_1=am/k_0 \) and \( t_2=bm/k_0 \) respectively. Based on these times, the reliability of the measurement apparatus $R$ can be assessed, as shown in Fig.~\ref{fig:reliability}. Systematic error is defined as \(\delta B_x = \tilde{B}_x - B_x\). When reliability is high and errors are minimal, $1- \text{reliability}$ exhibits a similar functional trend as the error, as shown in Fig.~\ref{fig:error}. In this regime, reliability can accurately characterize the systematic error. 

To determine the region where reliability is applicable, specifically where it can characterize systematic error, we analyze the derivative of \(\delta B_x\),
\begin{equation}
    \frac{\rm d}{{\rm d}B_x}\delta B_x = \frac{k_0 \hbar}{a m}\sqrt{\frac{\tilde \beta}{\tilde \alpha}} \frac{\left(\tilde{\alpha}\tilde{\beta}'-\tilde{\beta}\tilde{\alpha}'
    \right)}{2\tilde{\beta}(\tilde{\beta}+\tilde{\alpha})}-1.
\end{equation}
As \( B_x \to 0 \) and \( b \to \infty \), we expand the first term of the above equation and retain terms up to the first order
\begin{equation}
    \sqrt{\frac{\tilde \beta}{\tilde \alpha}} \frac{\left(\tilde{\alpha}\tilde{\beta}'-\tilde{\beta}\tilde{\alpha}'
    \right)}{2\tilde{\beta}(\tilde{\beta}+\tilde{\alpha})}
    \sim \frac{B_x}{\sqrt{{\rm Erf}(b/b_0) +B_x^2 }}.
\end{equation}
For a given \( b \), the minimum magnetic field at which reliability is applicable is $B_{\text{min}} \sim \sqrt{{\rm Erf}(b/b_0)}$, where $b_0=\sqrt{2}\hbar k_0/f m \sigma'$ is the characterized length of the system. When \( b \) is sufficiently large, \( B_{\text{min}} \to 0\) can be considered negligible.

\begin{figure}[htbp]
    \centering
    \includegraphics[width=0.95\linewidth]{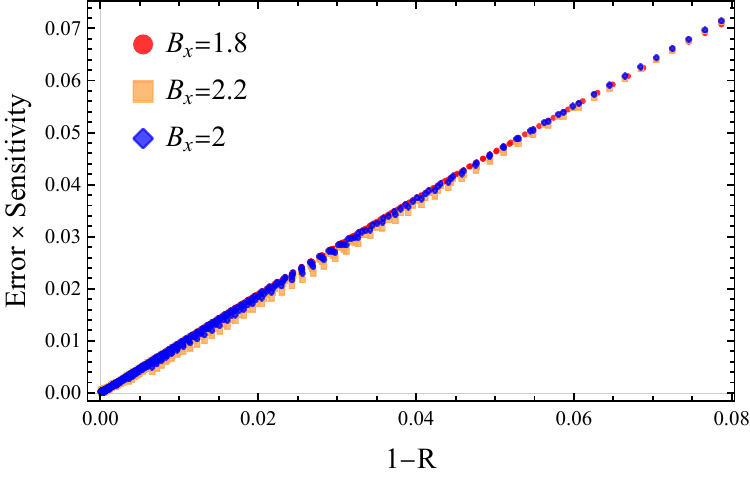}
    \caption{When the apparatus is near-ideal, the relationship among \(1-\text{reliability}\), sensitivity, and error is governed by Eq.~\eqref{eq:R and S and E}.Parameters: $b\in [30,50], \ k_0 \in [10,15]$.}
    \label{fig:validation of R S and E}
\end{figure}

In the end, we validate the relationship derived in Eq.~\eqref{eq:R and S and E}. When the system is nearly ideal, the relationship among reliability, sensitivity, and error is observed to be linear, as illustrated in Fig.~\ref{fig:validation of R S and E}. Here, it can be observed that the proportionality coefficient is independent of the magnetic field being measured. 

\section{Conclusion and discussion}
In this paper, we propose a general relation among reliability, sensitivity and error of measurement apparatus at first, as shown in Eq.~\eqref{eq:R and S and E}. Then, with the definition of quantum reliability, as outlined in ref.\cite{QuantumReliability}, we demonstrate this relationship through a specific quantum sensing model. We consider a specific quantum sensing process in which a spin is employed as a sensor to detect a magnetic field. The spin state is then read out using a Stern-Gerlach apparatus, allowing us to infer the magnitude of the magnetic field. In this process, spins scatter after passing through the magnetic field to be measured, and the SG apparatus may not perfectly separate the spin wave packets, leading to systematic errors. Through the analysis of quantum reliability, we verify the above argument,  thus, quantum reliability could be used to characterize the systematic error of the quantum sensing process. Additionally, we provide the parameter range within which this characterization is applicable.

\begin{figure}[htbp]
    \centering
    \includegraphics[width=0.95\linewidth]{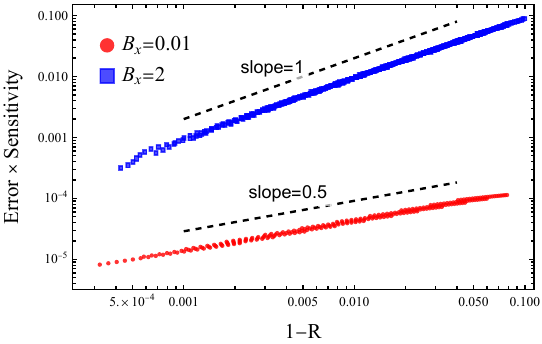}
    \caption{The scaling behavior differs when either the first-order or second-order terms dominate in Eq.~\eqref{eq:M_expansion}. The blue line represents the scenario where the first-order terms prevail, adhering to the relationship described in Eq.~\eqref{eq:R and S and E}. In contrast, the red line illustrates the case where the second-order terms are dominant. The sensitivity approaches zero, and \( \delta Q \sim (1-R)^{1/2} \). Parameters: $b\in [30,50], \ k_0 \in [13,15]$.}
    \label{fig:differentScaling}
\end{figure}

In deriving the relationship expressed in Eq.~\eqref{eq:R and S and E}, we considered only the first-order term in the expansion of Eq.~\eqref{eq:M_expansion}. However, when the first-order term is not dominant, Eq.~\eqref{eq:R and S and E} no longer holds. For example, in this particular case, as \(B_x\) approaches zero, \( {dM}/{dB_x} \to 0\). As a result, the second-order term \( {d^2M}/{dB_x^2} \times \delta^{2} B_x\)  is dominant. This implies the linear relationship between \(1-R\) and \(\text{Error} \times \text{Sensitivity}\) breaks down, and is replaced by \(\text{Error} \sim (1-R)^{1/2} \), as illustrated in Fig.~\ref{fig:differentScaling}.
Besides, for apparatuses reliant on highly non-linear behavior, such as criticality-enhanced quantum sensors \cite{Loschmidt_CPSun_2006,Quantum-criticality-as-a-resource-for-quantum-estimation,quantum-chaotic-sensors,Quantum_Critical_Metrology,Quantum_critical_detector}, whether this relationship holds requires further study. Additional investigations are required to assess the reliability of these systems. Another pertinent issue in quantum reliability, warranting further study, is the inverse problem addressed in this paper: specifically, the generation of suitable projection operators that can effectively characterize a quantum apparatus's operational efficacy.


This work opens a new avenue for the reliability analysis of quantum systems, representing a significant application of quantum reliability. We have analyzed the reliability of quantum sensing process by translating classical metrics used to evaluate measurement apparatus, such as error, into the language of quantum reliability. In the absence of standard reference apparatus, this approach allows us to assess the accuracy of the apparatus through their reliability.


\section{Acknowledgments}
The authors appreciate Hui Dong for his fruitful discussions. This work was supported by the National Natural Science Foundation of China (NSFC) (Grant No. 12088101) and NSAF No.U2330401.


\bibliography{JRSE}

\appendix
\label{Appendix:overlap integral}
\section{Equivalence between the overlap integral in SG experiment and the reliability projection in Eq.~\eqref{Eq:SG Realibality projection}}
For a Stern-Gerlach apparatus, an inhomogeneous magnetic field spatially separates spins in different directions. Denote the two resulting wave packets as $\psi^+(z)$ and $\psi^-(z)$. Here, we do not require the specific form of the Hamiltonian; we begin by writing out the overlap integral of the two wave packets:
\begin{equation}
    I_{\text {overlap}}=\big|\int_{-\infty}^\infty \psi^{+*}(z) \psi^-(z) \big| \ {\rm d}z.
\end{equation}
The reliability calculated through $E_2$ defined in Eq.~\eqref{Eq:SG Realibality projection} is
\begin{equation}
    R_{\text {SG}} = 
        \int_0^{\infty} |\psi^+(z) |^2  \ {\rm d}z +\int_{-\infty}^0 |\psi^-(z) |^2  \ {\rm d}z.
\end{equation}
Considering a near-ideal scenario where the up and down wave packets are almost completely separated, we can extend the limits of integration in \( R_{\text{SG}} \) to encompass the entire space. Applying the Cauchy-Schwarz inequality yields:
\begin{equation}
    I_{\text {overlap}} \leq R_{\text {SG}}.
\end{equation}
Considering that prolonged operation of the SG apparatus results in greater spatial separation of the two wave packets, we recognize that the overlap integral \( I_{\text {overlap}} \) and the reliability \( R_{\text{SG}} \) exhibit the same monotonic behavior. With the aforementioned inequality, we conclude that using \( R_{\text{SG}} \) to characterize the performance of the SG apparatus is equivalent to describing it in terms of the overlap integral.

\section{1D scattering in magnetic field}
\label{Appendix:1d scattering}
This appendix provides a detailed derivation of Eq. \eqref{eq:T and R} and Eq. \eqref{eq:U1Psi0}.

\begin{figure}[h]
    \centering
    \includegraphics[width=0.8\linewidth]{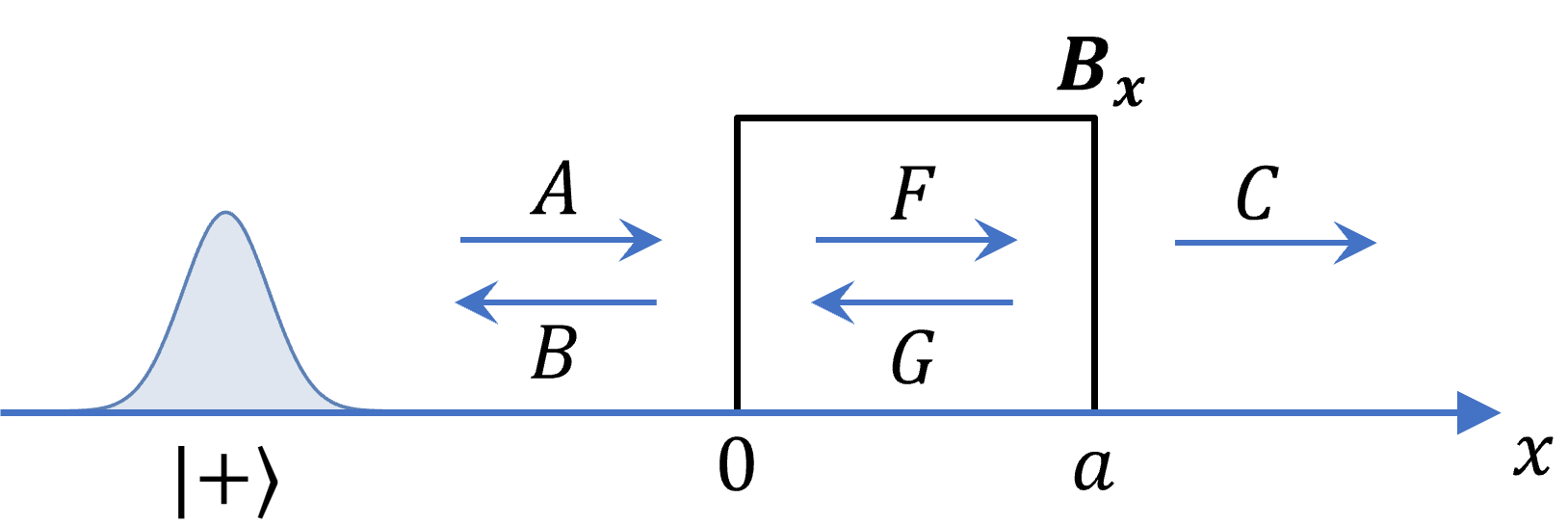}
    \caption{1d far-filed scattering.}
    \label{fig:1d far-field scattering}
\end{figure}

Taking the potential barrier$(+)$ as an example (Fig~\ref{fig:1d far-field scattering}). For Hamiltonian $H^{(+)}$, we first solve its scattering state $\phi_k^{(+)} \equiv \braket{x}{k^{(+)}}$, where $H^{(+)}\ket{k^{(+)}}=E_k \ket{k^{(+)}}$. Based on the Hamiltonians in different regions, the piecewise wave function can be expressed as:
\begin{itemize}
    \item For \( x < 0 \), set \(\phi_k^{(+)}(x) = A e^{ikx} + B e^{-ikx}\), where \( k = \sqrt{2mE_k/\hbar^2} \);
    \item For \( 0 < x < a \), set \(\phi_k^{(+)}(x) = F e^{iq_{(+)}x} + G e^{-iq_{(+)}x}\), where \( q_{(+)} = \sqrt{2m(E_k - B_x)/\hbar^2} \);
    \item For \( x > a \), set \(\phi_k^{(+)}(x) = C e^{ikx} \).
\end{itemize}
According to the continuity conditions, \ie, the continuity of the wave function and its derivative at \( x = 0 \) and \( x = a \), we obtain the following four conditions:
\begin{equation}
    \begin{aligned}
        T^{(+)}_k \equiv \frac{C}{A}=&\frac{4kq_{(+)}e^{i(q_{(+)}-k)a}}{(k+q_{(+)})^2-(k-q_{(+)})^2e^{2iq_{(+)}a}},\\
        R^{(+)}_k \equiv \frac{B}{A}=&\frac{(k^2 -q_{(+)}^2)(1-e^{2iq_{(+)}a})}{(k+q_{(+)})^2-(k-q_{(+)})^2 e^{2iq_{(+)}a}},\\
        M_{1k}^{(+)} \equiv \frac{F}{A}=&\frac{2k(k+q_{(+)})}{(k+q_{(+)})^2-(k-q_{(+)})^2 e^{2iq_{(+)}a}},\\
        M_{1k}^{(+)} \equiv \frac{F}{A}=&\frac{2k(k+q_{(+)})}{(k+q_{(+)})^2-(k-q_{(+)})^2 e^{2iq_{(+)}a}},\\
        M_{2k}^{(+)} \equiv \frac{G}{A}=&\frac{-2k(k-q_{(+)})e^{2iq_{(+)}a}}{(k+q_{(+)})^2-(k-q_{(+)})^2 e^{2iq_{(+)}a}}.\\
    \end{aligned}
\end{equation}
Thus, the scattering state can be obtained as:
\begin{equation}
    \phi_k^{(+)}(x) = 
    \begin{cases} 
        e^{ikx} + R^{(+)}_k e^{-ikx}, & x < 0 \\
        M_{1k}^{(+)} e^{iq_{(+)}x} + M_{2k}^{(+)} e^{-iq_{(+)}x}, & 0 < x < a \\
        T^{(+)}_k e^{ikx}, & x > a 
    \end{cases}.
    \label{eq:scattering states}
\end{equation}
Here, \(T^{(+)}_k\) and \(R^{(+)}_k\) are referred to as the transmission coefficient and reflection coefficient, respectively. It is easy to check that \(|T^{(+)}_k|^2 + |R^{(+)}_k|^2 = 1\).

Next, we decompose the initial state $\ket{\psi_{x0}}$ in terms of the scattering states $\ket{k^{(+)}}$ and demonstrate that the expansion coefficients in the scattering state basis are equivalent to those in the plane wave basis. 

The expansion through plane wave reads
\begin{equation}
        \psi_{x0}(x)=\frac{1}{\sqrt{2\pi}}\int_{-\infty}^{\infty} \varphi(k) e^{ikx} \, {\rm d}k.
\end{equation}
The expansion coefficients are
\begin{equation}
    \begin{aligned}
        \varphi(k)&=\frac{1}{\sqrt{2\pi}} \int_{-\infty}^{\infty} \psi_{x0}(x) e^{-ikx} \, {\rm d}x\\
        &=\left(\frac{2\sigma^2}{\pi}\right)^{1/4} \exp\left[-ikx_0 - \sigma^2 (k-k_0) \right].
    \end{aligned}
\end{equation}
The expansion in the scattering state reads
\begin{equation}
        \ket{\psi_{x0}}
        = \frac{1}{\sqrt{2\pi}} \int_{-\infty}^{\infty} \tilde{\varphi}(k) \ket{k^{(+)}} {\rm d}k,
\end{equation}
where the expansion coefficients are
\begin{widetext}
\begin{equation}
    \begin{aligned}
        \tilde{\varphi}(k) 
        &\equiv  \frac{1}{\sqrt{2\pi}}  \int_{-\infty}^{\infty} \braket{k^{(+)}}{x} \braket{x}{\psi_{x0}}  {\rm d} x
        = \frac{1}{\sqrt{2\pi}}  \int_{-\infty}^{\infty} \phi_k^{(+)*}(x) \psi_{x0}(x) \, {\rm d} x\\
        &= \frac{1}{\sqrt{2\pi}}  \int_{-\infty}^{0} (e^{-ikx}+R_k^{(+)*} e^{ikx})  {\rm d} x 
        + \frac{1}{\sqrt{2\pi}}  \int_{0}^{a} (M_{1k}^{(+)*} e^{-iq_{(+)}x}+M_{2k}^{(+)*} e^{iq_{(+)}x}) \, {\rm d} x 
        +\frac{1}{\sqrt{2\pi}}  \int_{a}^{\infty} (T_k^{(+)*} e^{-ikx})  {\rm d} x \\
        &\approx  \frac{1}{\sqrt{2\pi}}  \int_{-\infty}^{\infty} (e^{-ikx}+R_k^{(+)*} e^{ikx})  {\rm d} x 
        \approx \varphi(k) + R_{k_0}^{(+)*} \varphi(-k) \approx \varphi(k).
    \end{aligned}
\end{equation}
\end{widetext}
The approximate equality arises because the initial wave packet is localized far to the left, and its momentum is concentrated around \( k_0 > 0 \). At this point, the initial state can be expressed as an expansion in terms of the scattering states
\begin{equation}
    \psi_{x0}(x)=\int_{-\infty}^{\infty} \varphi(k) \phi_k^{(+)}(x) \, {\rm d}k.
\end{equation}

We now calculate the time evolution of the wave function to determine the outgoing state. The wave function at time $t$ reads
\begin{equation}
    \psi_{x0}^{(+)}(x,t)=\int_{-\infty}^{\infty}  \varphi(k) \phi_k^{(+)}(x) e^{-i\frac{\hbar k^2}{2m} t}  \, {\rm d}k
\end{equation}
Substituting the scattering state from Eq.~\eqref{eq:scattering states} into the above expression, for \( t \gg 0 \), the parameter \( e^{-i\frac{\hbar k^2}{2m} t} \) undergoes rapid oscillations, leading to the outgoing state given by
\begin{equation}
    \psi_{x0}^{(+)}(x,t) = \begin{cases}
        R^{(+)} \psi_{x0,-}(x,t), & x<0\\
        T^{(+)} \psi_{x0,+}(x,t), & a<x\\
    \end{cases}
\end{equation}
where 
\begin{equation}
    \psi_{x0,\pm}(x,t) = \int_{-\infty}^{\infty}  \varphi(k) \exp\left[\pm ikx -i\frac{\hbar k^2}{2m} t \right] {\rm d} k
\end{equation}
are the wave function transmission($+$) and reflection($-$), respectively. 

Applying the same method to the \(|-\rangle\) state, we obtain the total outgoing wave function in the magnetic field region as
\begin{equation}
\begin{aligned}
    &U_1 \ket{\Psi_0} =  \\
    &\begin{cases}
    \frac{1}{\sqrt{2}} \left(R^{(+)} \ket{+} + R^{(-)} \ket{-} \right)\otimes \ket{\psi_{x0,-}}\otimes \ket{\psi_{z1}}, & x<0\\
    \frac{1}{\sqrt{2}} \left(T^{(+)} \ket{+} + T^{(-)} \ket{-} \right)\otimes \ket{\psi_{x0,+}}\otimes \ket{\psi_{z1}}, & a<x\\
\end{cases}.
\end{aligned}
\end{equation}
Here, the wave function in the z-direction has propagated for a time \( t_1 \), resulting in a Gaussian wave packet $\psi_{z1}$ with a width of \(\sigma' = \sigma^2(1+i\hbar t_1 /m\sigma^2 )\).

\end{document}